\documentclass[sigconf]{acmart} 
\usepackage{pifont}% http://ctan.org/pkg/pifont
\usepackage{multirow}
\usepackage{multicol}
\usepackage{subfigure}
\usepackage{bm}
\usepackage{color}

\usepackage{enumitem}
\usepackage{bbm}
\usepackage{graphicx}
\usepackage{mathtools}
\usepackage[linesnumbered,ruled,vlined]{algorithm2e}
\usepackage{bbm}
\usepackage{balance}
\usepackage{graphics}
\usepackage{makecell}
\AtBeginDocument{%
  \providecommand\BibTeX{{%
    \normalfont B\kern-0.5em{\scshape i\kern-0.25em b}\kern-0.8em\TeX}}}

\setcopyright{none}

\acmConference[Conference acronym 'XX]{Make sure to enter the correct
  conference title from your rights confirmation emai}{June 03--05,
  2018}{Woodstock, NY}

\usepackage{stfloats,acmart-taps}
\acmSubmissionID{39}
\begin{document}

\title{Equivariant Contrastive Learning for Sequential Recommendation}

\author{Peilin Zhou}
\affiliation{%
  \institution{The Hong Kong University of Science and Technology (Guangzhou)}
  \country{China}}
  \email{zhoupalin@gmail.com}

\author{Jingqi Gao}
\affiliation{%
  \institution{Upstage}
  \country{Hong Kong}
}
\email{mrgao.ary@gmail.com}

\author{Yueqi Xie}
\affiliation{%
  \institution{The Hong Kong University of Science and Technology}
  \country{Hong Kong}}
\email{yxieay@connect.ust.hk}

\author{Qichen Ye}
\affiliation{%
  \institution{Peking University}
  \country{China}}
  \email{yeeeqichen@pku.edu.cn}

\author{Yining Hua}
\affiliation{%
  \institution{Harvard University}
  \country{USA}
}
\email{yininghua@g.harvard.edu}

\author{Jae Boum Kim}
\affiliation{%
  \institution{The Hong Kong University of Science and Technology}
  \country{Hong Kong}}
  \email{jbkim@cse.ust.hk}

\author{Shoujin Wang}
\affiliation{%
  \institution{University of Technology Sydney}
  \country{Australia}}
  \email{shoujin.wang@uts.edu.au}
  
\author{Sunghun Kim}
\affiliation{%
  \institution{The Hong Kong University of Science and Technology (Guangzhou)}
  \country{China}}
  \email{hunkim@cse.ust.hk}

\renewcommand{\shortauthors}{P. Zhou et al.}
\renewcommand{\shorttitle}{Equivariant Contrastive Learning for Sequential Recommendation}

\begin{abstract}
Contrastive learning (CL) benefits the training of sequential recommendation models with informative self-supervision signals. Existing solutions apply general sequential data augmentation strategies to generate positive pairs and encourage their representations to be invariant.
However, due to the inherent properties of user behavior sequences, some augmentation strategies, such as item substitution, can lead to changes in user intent. Learning indiscriminately invariant representations for all augmentation strategies might be sub-optimal.
Therefore, we propose \textbf{E}quivariant \textbf{C}ontrastive \textbf{L}earning for \textbf{S}equential \textbf{R}ecommendation (ECL-SR), which endows SR models with great discriminative power, making the learned user behavior representations sensitive to invasive augmentations (e.g., item substitution) and insensitive to mild augmentations (e.g., feature-level dropout masking).
In detail, we use the conditional discriminator to capture differences in behavior due to item substitution, which encourages the user behavior encoder to be equivariant to invasive augmentations.
Comprehensive experiments on four benchmark datasets show that the proposed ECL-SR framework achieves competitive performance compared to state-of-the-art SR models.
The source code is available at \url{https://github.com/Tokkiu/ECL}.
\end{abstract}





\maketitle
\begin{figure}[t]
  \centering
  \includegraphics[width=85mm]{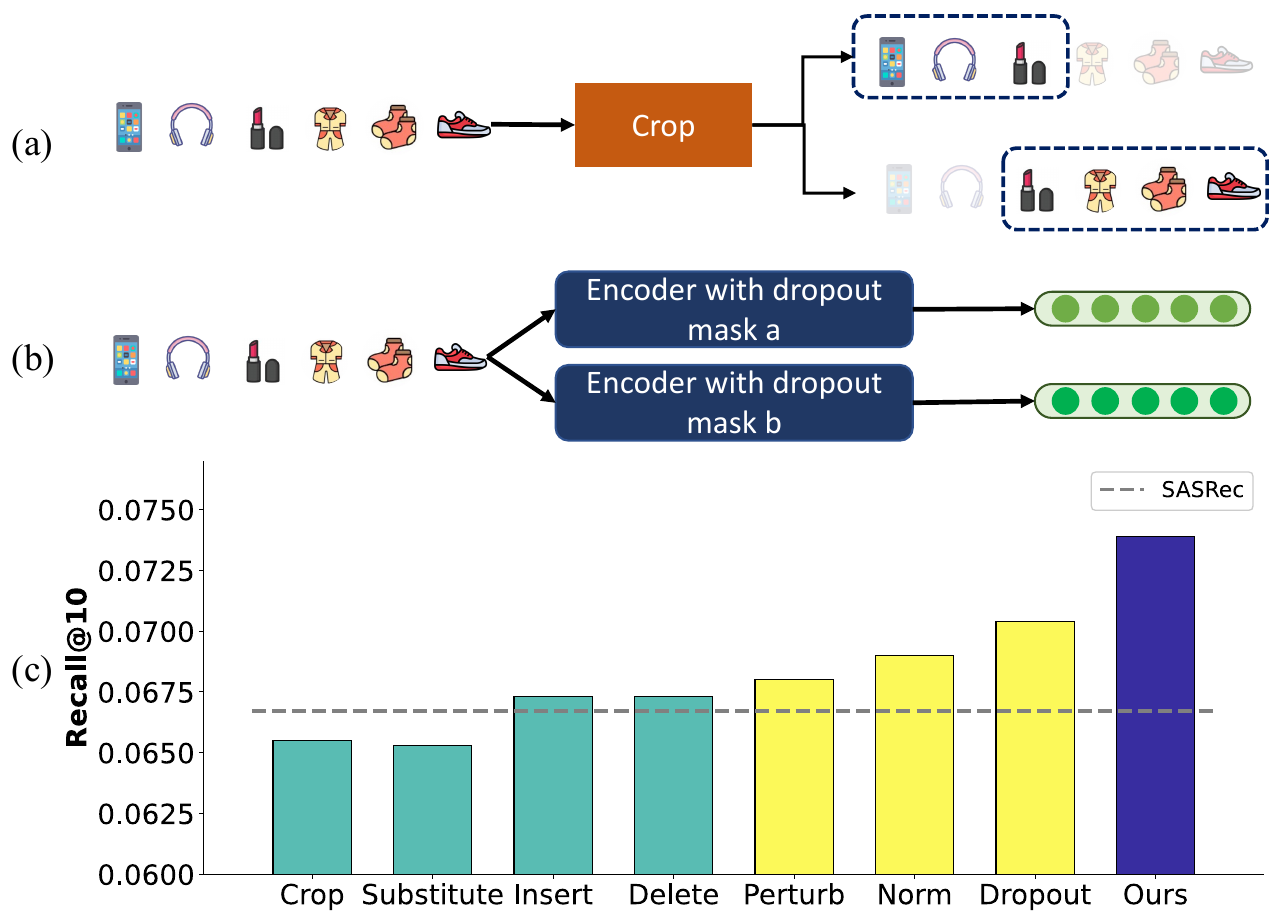}
  \caption{(a) An example of invasive data augmentation; (b) An example of mild augmentation; (c) Performance comparison of using different data augmentations to generate positive pairs for invariant contrastive learning on Yelp dataset. }
\label{fig:intro}
\end{figure}
\section{Introduction}
\label{sec-intro}
Sequential recommendation aims to predict the next item which may interest a given user via learning the user's dynamic preference from her/his interaction sequence with items ~\cite{wang2019sequential,song2021next}. As an important recommendation paradigm, sequential recommendation has been playing a vital role in various Web service domains such as e-commerce~\cite{hwangbo2018recommendation,wang2020era}, social media~\cite{guy2010social}, video site~\cite{davidson2010youtube}, etc. 
    Despite extensive research and significant progress in recent years, sequential recommendation still faces a significant challenge: data sparsity. This issue arises because that the user-item interaction data, which is the foundation of sequential recommendation, is typically limited compared with a large number of users and items, i.e., millions of users or items. To address this challenge, self-supervised learning (SSL) has attracted increasing attention in recent years due to its powerful capability to alleviate data sparsity issues by mining self-supervised signals from raw user-item interaction data~\cite{ssr_survey,Hyperbolic}. Consequently, various studies have been done to develop more accurate sequential recommender systems by incorporating SSL into sequential recommendation in recent years. These studies generally focus on exploring various data augmentation strategies to enrich and enhance the input data of recommender systems, thereby  improving their recommendation performance. For example, S$^{3}$Rec~\cite{zhou2020s3} is the first work to utilize item masking and cropping techniques to augment users' sequential interaction data and design corresponding pretext tasks for model pretraining for sequential recommendations. CL4SRec~\cite{cls4rec} applies three sequence-based operations for data augmentation to improve the performance of sequential recommendation: item masking, sequence reordering, and sequence cropping. DuoRec~\cite{duorec} further combines both unsupervised and supervised data augmentation methods to mitigate representation degeneration in sequential recommendation.

According to the augmented object, existing augmentation strategies can be divided into two categories: sequence-level augmentation (e.g., item cropping shown in Fig.~\ref{fig:intro} (a)) and feature-level augmentation (e.g., dropout shown in Fig.~\ref{fig:intro} (b)). The former directly operates on the user-item interaction sequences, while the latter operates in the latent feature space. Intuitively, augmentations at the sequence level are more likely to cause significant semantic shifts, namely, leading to an unexpected deviation from the original user behavior, so we define them as "\textbf{invasive}" augmentations. In contrast, the impact of augmentations at the feature level on semantics is easier to control compared to that of sequence-level augmentations, and thus feature-level augmentations are considered as "\textbf{mild}".   
These two categories of augmentations are indiscriminately utilized in existing contrastive learning frameworks~\cite{cls4rec, duorec}, which encourage the learned user behavior representations to be invariant to the variations caused by augmentation strategies. Here, "invariant" means the representaion learned from the original instance and that learned from its corresponding augmented instance via contrastive learning are similar. 
However, the rationale of this invariant contrastive learning paradigm for sequential recommendation still needs to be further examined. In fact, we observe that different positive instances produced from the same original instance using invasive augmentation strategies may not be semantically "identical". The main reason is that invasive augmentation methods, such as item cropping, insertion, and substitution, may break critical associations among items existing in original user-item interaction sequences. For example, as shown in Fig.~\ref{fig:intro} (a), when we apply random item cropping to a user behavior sequence (i.e., "\textit{smartphone, headset, lipstick, blouse, socks, sneakers}"), the two generated positive instances ("\textit{smartphone, headset, lipstick}" and "\textit{lipstick, blouse, socks, sneakers}") do not share identical semantics. The former primarily focuses on digital products, whereas the latter mainly focuses on clothes. This problem may become even worse for short interaction sequences due to their greater vulnerability~\cite{coserec}.

To empirically verify the aforementioned observation, we investigate a typical Invariant Contrastive Learning (ICL) approach called CL4SRec~\cite{cls4rec}, which primarily aims to learn invariant representations for the positive instances generated by diverse augmentation strategies. Using CL4SRec as the backbone, we compare the performance of invariant contrastive learning with different sequence-level (invasive) augmentations and feature-level (mild) augmentations. A detailed introduction to the augmentations can be found in Sec.~\ref{sec-aug}. As demonstrated by Fig.~\ref{fig:intro} (c), CL4SRec~\cite{cls4rec} with additional feature-level augmentations (yellow bars) consistently outperforms the base model (indicated by the dashed line). However, when the feature-level augmentations employed in CL4SRec are replaced by sequence-level augmentations, the performance (green bars) is less satisfactory\footnote{Although insertion and deletion operations slightly enhance the performance, sequence-level augmentations are more prone to produce semantic drift compared to feature-level augmentations. Therefore, we still categorize them as invasive augmentations.}.
 Sequence-level augmentations cannot consistently improve the recommendation performance and sometimes even degrade it. The empirical study indicates that the current invariant contrastive learning paradigm is more suitable for mild augmentations conducted at the feature level. Consequently, a natural question
arises: \textit{\textbf{How can we build a more reliable contrastive learning framework that benefits from both mild feature-level augmentations and invasive sequence-level augmentations to further improve the performance of sequential recommendation?}} 

To bridge this significant gap, this paper proposes a novel framework called \textbf{E}quivariant \textbf{C}ontrastive \textbf{L}earning for \textbf{S}equential \textbf{R}ecommendation (ECL-SR). ECL-SR is able to build powerful contrastive learning based on both mild feature-level augmentations and invasive sequence-level augmentations to learn more informative representations for sequential recommendation. 
The core idea behind ECL-SR is to learn sequence representations that can identify, rather than being invariant to, the differences induced by invasive sequence-level augmentations, while preserving the invariance learning for mild augmentations.
\begin{figure}[t]
  \centering
  \includegraphics[width=85mm]{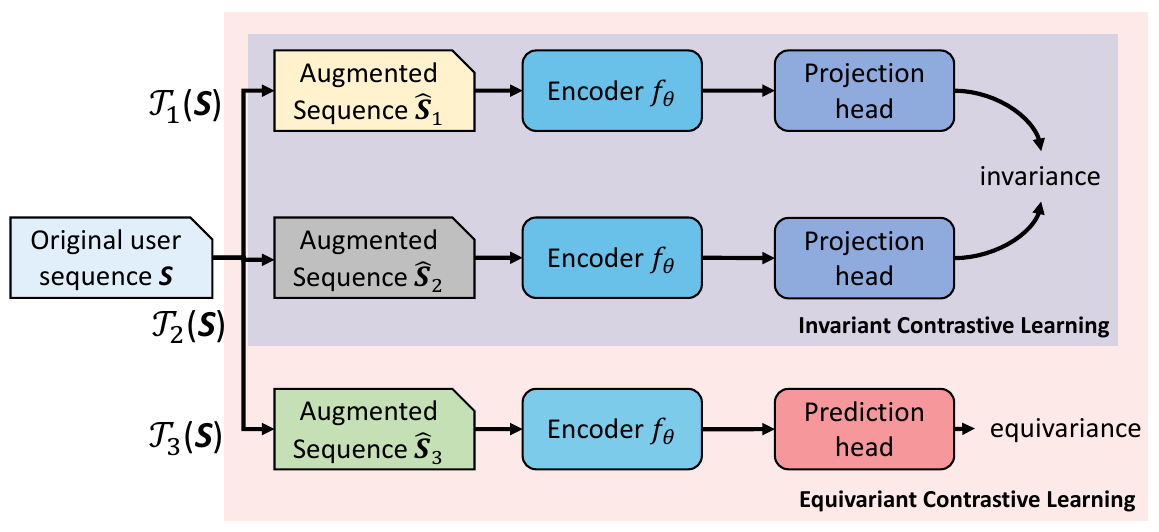}
  \caption{The connection between invariant contrastive learning (ICL) and equivariant contrastive learning (ECL): ECL is a generalization of traditional ICL methods that highlights the complementary nature of invariance and equivariance. }
\label{fig:preliminary}
\end{figure}
Mathematically, ECL-SR leverages mild and invasive augmentations within a unified framework to encourage both the \textit{invariance} property and the \textit{equivariance} property of contrastive learning for learning more informative representations. Invariance makes representations insensitive to non-essential variations, while equivariance encourages representations to change predictably in response to augmentations. Specifically, ECL-SR learns invariance and equivariance by adopting a contrastive loss, such as InfoNCE~\cite{oord2018representation} loss, on mild augmentations and a prediction loss on invasive augmentations, respectively.
Fig.~\ref{fig:preliminary} illustrates the connection between equivariant contrastive learning and invariant constrastive learning.
In fact, invariant constrastive learning can be regarded as a special case of equivariant contrastive learning, which we thoroughly discuss in Sec.\ref{ECL}.

Experiments on four benchmark datasets show that ECL-SR outperforms both basic SR models and invariant contrastive learning based SR models by effectively leveraging the complementary nature between mild augmentations and invasive augmentations with equivariant contrastive learning. Furthermore, we dive into the effectiveness of various augmentation strategies and the impact of components and hyperparameters in ECL-SR.

Our primary contributions can be summarized as follows:
\begin{itemize}
    \item We propose the ECL-SR framework,  which effectively utilizes both mild and invasive augmentation to enrich user behavior representations. 
    \item We further instantiate the ECL-SR framework by incorporating dropout as the mild augmentation and masked item substitution as the invasive augmentation, illustrating the synergy effect between these two types of augmentations. 
    \item We employ a generator-discriminator architecture to implement masked item substitution and capture user behavior discrepancies between original interaction sequences and their augmented counterparts, thererby facilitating the learning of equivariance for invasive augmentations.
    \item We conduct comprehensive experiments on 4 benchmark SR datasets, which demonstrate the advantages of ECL-SR over both classical SR models and state-of-the-art contrastive learning based SR models. 
\end{itemize}
\begin{table}[t]
\caption{Notation.}
\label{tb:notation}
 \resizebox{75mm}{!}{
\begin{tabular}{l l}
\toprule
Notation&Description\\
\midrule
$\mathcal{U},\mathcal{I}$        & user and item set\\
$v^u_{t}$ & the $t$-th item that user $u$ interacted\\
$\vmathbb{S}_u$ & chronological interaction sequence for a user\\
$\vmathbb{S}_u^{+}$ & \makecell[l]{positive view of the $\vmathbb{S}_u$} \\
$\vmathbb{S}_{u}^{\prime}$ & \makecell[l]{masked version of the $\vmathbb{S}_u$} \\
$\vmathbb{S}_{u}^{\prime \prime}$ & \makecell[l]{partially substituted version of the $\vmathbb{S}_u$} \\
$L$ & maximum sequence length\\
$\mathbf{h}_{t}^{u}$ & \makecell[l]{latent vector at $t$-th timestamp}  \\
$\mathbf{h}^{u}$ & \makecell[l]{ aggregated behavior representation for user $u$}  \\
$\mathcal{T}_*(\cdot)$ & data augmentation operator\\
$f_\theta$ & user behavior encoder (UBE)\\
$f_\mathcal{G}$ & generator (G)\\
$f_\mathcal{D}$ & conditional discriminator (CD)\\
$\phi_{\zeta}(\cdot)$ & prediction head for detecting invasive augmentations\\
$\alpha$, $\beta$, $\lambda$ & balancing hyper-parameters\\
$k$ & window size for aggregation\\
$\gamma$ & mask ratio\\
\bottomrule
\end{tabular}
}
\end{table}
\section{PRELIMINARIES}
\subsection{Problem Formulation}
\label{sec-problem}
We denote $\mathcal{U}$ and $\mathcal{I}$ as the user set and item set, respectively. 
For each user $u \in \mathcal{U}$, his/her chronological interaction sequence can be represented as $\vmathbb{S}_u=[v^u_1, v^u_2, \dots,v^u_{t}, \dots, v^u_{L}]$, where $v^u_{t}$ refers the $t$-th item that user $u$ interacted and $L$ is the maximum sequence length.
The goal of sequential recommendation is to predict the next item $v^u_{L+1}$ that user $u$ will most likely interact with given the interaction history $\vmathbb{S}_u$.
To achieve this goal, the typical training loss for SR is usually designed as follows:
\begin{equation}
\label{eq:rec_loss}
\mathcal{L}_{\mathrm{Rec}} = \sum_{u=1}^{|\mathcal{U}|} \sum_{t=2}^{L} -\log p_{\theta}(v^{u}_{t+1}|v^{u}_{1}, v^{u}_{2}, \cdots, v^{u}_{t}),
\end{equation}
where $\theta$ represents the parameters of a neural network $f_{\theta}$, which encodes sequential patterns into latent vectors:
$\mathbf{h}_{t}^{u} = f_{\theta}(\{v^{u}_{j}\}_{j=1}^{t})$. The probability $p(v^{u}_{t+1} \mid \mathbf{h}_{t}^{u})$ is calculated based on the similarity between the encoded sequential behaviors $h_{t}^{u}$
and the embedding of the next item $v^{u}_{t+1}$ in the representation space.
For inference, the items with top-N probability $p(v^{u}_{L+1} \mid \mathbf{h}_{L}^{u})$ will be recommended to the user $u$.

\subsection{Data Augmentation Strategies}
\label{sec-aug}
Given the original user sequence $\vmathbb{S}_u$, several random sequence-level (invastive) augmentation  strategies can be employed:
\begin{itemize}
    \item \textbf{Insertion}. It first randomly selects a position within $\vmathbb{S}_u$ and then inserts an item, chosen randomly from the interaction histories of other users, into that position. This strategy is performed multiple times on the sequence to generate an augmented version. The augmented example could be represented by:
    \begin{equation}
        \vmathbb{S}_u^i = [v_{1}^u, v_{in}^u, v_{2}^u, ..., v_L^u]
    \end{equation}
    \item \textbf{Deletion}. It randomly deletes an item in the original sequence and re-runs the operation to form an augmented sequence:
     \begin{equation}
        \vmathbb{S}_u^d = [v_{1}^u, v_{3}^u, ..., v_L^u]
    \end{equation}
    \item \textbf{Substitution}. It randomly selects a proportion of items from $\vmathbb{S}_u$ to be substituted as $l_r$. The items in $l_r$ are randomly selected from all negative samples for $\vmathbb{S}_u$. The substitution ratio is empirically set to 0.2. An example of the substituted sequence is shown below:
    \begin{equation}
        \vmathbb{S}_u^s = [v_{1}^u, v_{r}^u, v_{3}^u, ..., v_L^u]
    \end{equation}
    \item \textbf{Cropping}. It randomly selects a continuous sub-sequence from positions $i$ to $i+l_c$ from $\vmathbb{S}_u$ and removes it. The length to crop ($l_c$) is defined by $l_c = \alpha  |\vmathbb{S}_u|$ where empirically $\alpha = 0.8$. An example of the cropped sequence is shown below:
    \begin{equation}
        \vmathbb{S}_u^c = [v_{i}^u, v_{i+1}^u, ..., v_L^u]
    \end{equation}
    \item \textbf{Reordering}. It randomly selects a continuous sub-sequence from positions $i$ to $i+l_c$ from $\vmathbb{S}_u$ and shuffles it. The length of reordering ($l_c$) is defined by $l_c = \alpha  |\vmathbb{S}_u|$ where empirically $\alpha = 0.2$. An example of the reordered sequence is shown below:
    \begin{equation}
        \vmathbb{S}_u^c = [v_{1}^u, v_{3}^u,v_{4}^u, v_{2}^u, v_{5}^u, ..., v_L^u]
    \end{equation}
\end{itemize}

Given the user representation ${h^u}$, the following mild augmentation strategies can be applied:
\begin{itemize}
    \item \textbf{Perturbation}. It obtains the random noise according to the representation $h^u$ for augmentation. Formally, given the $h^u$ in a $d$-dimensional embedding space, the perturbation operation can be implemented in the following way:
    \begin{equation}
        h^u_{p}=h^u+\Delta_u
    \end{equation}
    The noise vectors $\Delta$ are subject to the following constraints:
    \begin{equation}
    \begin{split}
        &||\Delta||_2 = \epsilon, \\
        &\Delta = \hat{\Delta} \odot h^u, \hat{\Delta} \in \mathcal{R}^d \sim{U(0,1)},
    \end{split}
    \end{equation}
    As elaborated in SimGCL~\cite{perturb}, the constraints contribute to controlling the magnitude of $\Delta$ and the deviation of $h^u$, which help to retain most of the information from the original representation while still maintaining some variance. It is worth noting that the random noise added to each presentation is different.
    \item \textbf{Normalization}. It directly applies normalization operation on $h^u$. The operation preserves most of the information of the original representation while also adjusting the entire representation space in a gentle way to generate positive samples. Meanwhile, it helps alleviate popularity bias, as demonstrated in previous NISER~\cite{norm}. The operation can be implemented as follows:
    \begin{equation}
        h^u_{n} = \frac{h^u}{||h^u||_2}
    \end{equation}
\end{itemize}

\subsection{Invariant Contrastive Learning for Sequential Recommendation}
\label{sec-icl-sr}
In this section, we describe how existing works~\cite{cls4rec,duorec,coserec} apply invariant contrastive learning to sequential recommendation.
The fundamental idea behind these methods is to introduce an auxiliary task as well as a CL loss (e.g., InfoNCE~\cite{oord2018representation} loss) to help mine self-supervision signals. 
Specifically, as shown in Fig.\ref{fig:preliminary}, 
different data augmentation methods (either invasive or mild as described in Sec.~\ref{sec-intro}) are applied to the original sequence to generate positive views~\footnote{For brevity, we only show invasive augmentations in Fig.~\ref{fig:preliminary}. In practice, mild augmentations, such as dropout masking, can also be used.}.
Correspondingly, views from different sequences are considered as negatives.
Afterward, the CL loss is used to pull positive views closer and push negative views apart in the embedding space.
This essentially encourages the user sequence encoder to be insensitive to various data augmentation methods, leading to a more generalizable user behavior representation.
The recommendation task and the auxiliary task are usually trained jointly as follows:
\begin{equation}
\label{eq:cl_jl_loss}
\begin{aligned}
\theta^*=\underset{\theta}{\arg \min }\left[\mathcal{L}_{Rec}\left(f_\theta(\vmathbb{S}_u)\right)+\alpha \mathcal{L}_{ICL}\left(f_\theta(\hat{\vmathbb{S}}^{1}_u),f_\theta(\hat{\vmathbb{S}}^{2}_u)\right)\right],
\end{aligned}
\end{equation}
where $\mathcal{L}_{Rec}\left(f_\theta(\vmathbb{S}_u)\right)$ is the recommendation loss that has already been described in Sec.~\ref{sec-problem} and $\mathcal{L}_{ICL}\left(f_\theta(\hat{\vmathbb{S}}^{1}_u),f_\theta(\hat{\vmathbb{S}}^{2}_u)\right)$ represents the contrastive loss, which usually adopts InfoNCE loss that maximizes the mutual information (MI) between two positive views $\hat{\vmathbb{S}}^{1}_u \sim \mathcal{T}_1(\mathcal{\vmathbb{S}}_u)$ and $\hat{\vmathbb{S}}^{2}_u \sim \mathcal{T}_2(\mathcal{\vmathbb{S}}_u)$, $\mathcal{T}_1(\cdot)$ and $\mathcal{T}_2(\cdot)$ denote augmentation methods sampled from invasive augmentation set like \{crop, insert, deletion, reorder, ...\} or mild augmentation set like \{dropout, norm, perturb, ...\}; $\alpha$ is a hyper-parameter that balances the magnitude of these two losses. 

\begin{figure*}[t]
  \centering
  \includegraphics[width=140mm]{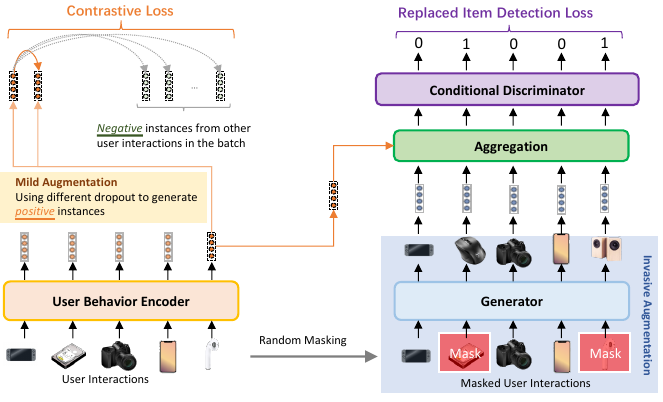}
  \caption{Overview of the proposed ECL-SR framework.}
\label{fig:overview}
\end{figure*}
\section{Methodology}
\subsection{Equivariant Contrastive Learning for Sequential Recommendation}
\label{ECL}
As discussed in Sec.~\ref{sec-intro} and Sec.~\ref{sec-icl-sr}, invariant contrastive learning based SR methods encourage representations that are all insensitive to user sequence augmentation methods. 
A key prerequisite for their effectiveness is that the chosen augmentation method introduces only non-essential changes in the original sequence without altering the semantics~\cite{coserec}.
However, some invasive augmentation methods, such as random crop and substitution, are prone to violate this premise and consequently impact the effectiveness of invariant contrastive learning for SR.
In this paper, we propose to train a neural network to sensitively detect the differences caused by invasive augmentations.
To achieve this, we generalize invariant contrastive learning to equivariant contrastive learning (ECL) for SR. We show the high-level structure of the framework in Fig.~\ref{fig:preliminary}. The upper part is used to learn invariance for mild augmentations and the lower part is designed to predict invasive augmentations for learning equivariance.

Following~\cite{ecl, diffcse}, the concept of equivariance can be defined as
 $f_\theta\left(\mathcal{T}_g(\boldsymbol{\vmathbb{S}_u})\right)=\mathcal{T}_g^{\prime}(f_\theta(\boldsymbol{\vmathbb{S}_u}))$, where $g \in A$ and $A$ is the group of invasive data augmentation methods, $\mathcal{T}_g(\vmathbb{S}_u)$ denotes the function with which $g$ augments an input user interaction sequence $\vmathbb{S}_u$, $f_\theta$ is the user behavior encoder that encode dynamic user interest into behavior representation $f_\theta(\vmathbb{S}_u)$, and $\mathcal{T}_g^{\prime}$ denotes a fixed transformation\footnote{In ECL-SR, the transformation $\mathcal{T}_g^{\prime}$ is obtained implicitly through the RIDL training process, which is described in Sec.~\ref{invasive}.}. 
It is worth noting that the ICL method is essentially a subcase of ECL-SR, where the identity function is used for $\mathcal{T}_g^{\prime}$ so that $f_\theta\left(\mathcal{T}_g(\vmathbb{S}_u)\right)=f_\theta(\vmathbb{S}_u)$. 

Finally, the optimization objective of ECL-SR for a user sequence $\vmathbb{S}_u$ is as follows\footnote{We omit the notion of head parameters in Equation 2 and Equation 3 for simplicity, but in practice, these parameters are used as part of the model and are updated during model training.}:
\begin{equation}
\begin{aligned}
\label{eq:ecl_jl_loss}
\theta^*=\underset{\theta}{\arg \min }[\mathcal{L}_{Rec}(f_\theta(\vmathbb{S}_u))&+\alpha \mathcal{L}_{ICL}(f_\theta(\hat{\vmathbb{S}}^{1}_u), f_\theta(\hat{\vmathbb{S}}^{2}_u)) \\
&+\beta \mathcal{L}_{ECL}(g, \phi_{\zeta}(f_\theta(\mathcal{T}_g(\vmathbb{S}_u)))],
\end{aligned}
\end{equation}
where $\hat{\vmathbb{S}}^{1}_u$ and $\hat{\vmathbb{S}}^{2}_u$ are two positive views generated by using mild augmentation methods; $g$ is sampled from invasive augmentation methods;
$\phi_{\zeta}(\cdot)$ denotes the prediction head for identifying the invasive augmentations;
$\alpha$ and $\beta$ are balancing hyper-parameters.
The goal of $\mathcal{L}_{ECL}$ is to predict the invasive augmentation $g$ using the prediction head output $\phi_{\zeta}(f_\theta(\mathcal{T}_g(\vmathbb{S}_u)))$, which encourages the shared encoder be equivariant to the invasive augmentations.

In the following sections, we will provide a detailed explanation of the ECL-SR framework and how it can be implemented.
Fig.~\ref{fig:overview} shows the overall structure of instantiated ECL-SR, which consists of three main components:
a User Behavior Encoder (UBE), a Generator (G), and a Conditional Discriminator (CD).
The primary objective of the UBE is to capture key patterns from user interactions and recommend the most suitable item to the user. 
In \textbf{Sec.~\ref{mild}}, we introduce ICL to the UBE, which helps learn invariant features for mild augmentation.
Invasive augmentation is encouraged using G and CD together to learn equivariance, as explained in \textbf{Sec.~\ref{invasive}}.
In the implementation, we instantiate the UBE and CD with SASRec~\cite{sasrec} (denoted as $f_{\theta}(\cdot)$ and $f_{\mathcal{D}}(\cdot)$ respectively), the G with BERT4Rec~\cite{bert4rec} (denoted as $f_{\mathcal{G}}(\cdot)$), which consists of several stacked self-attention blocks.
We choose dropout masking and item substitution as examples of mild and invasive augmentation, respectively, based on previous studies~\cite{duorec}.
We also conduct comprehensive experiments in Sec.~\ref{rq2} to analyze more augmentation combinations in the proposed framework.
Our notation is summarized in Tab. \ref{tb:notation}.

\subsection{Learning Invariance for Mild Augmentation}
\label{mild}
To encourage representations generated from the UBE to be insensitive to the mild augmentation, we employ invariant contrastive learning for the UBE, as shown in the left branch of Fig.~\ref{fig:overview}. Specifically, we generate the positive instance $\vmathbb{S}_u^{+}$ by applying feature-level dropout masking as the mild augmentation by default.  Other samples in the batch are regarded as negative instances. 

First, with the latent representations $\mathbf{h}_{t}^{u} = f_{\theta}(\{v^{u}_{j}\}_{j=1}^{t})$ of user behavior sequence $\vmathbb{S}_u$, inspired by~\cite{duorec,simcse,Yuan2021DualSA}, 
we take the average of the last $k$ representations to obtain an aggregated representation $\mathbf{h}_{u}$ with window size $k$ as follows\footnote{For brevity, we only show the situation when $k = 1$ in Fig~\ref{fig:overview}. Readers can refer to Sec.~\ref{rq3} for the impact of different values of $k$.}: 
\begin{equation}
\mathbf{h}_{u} = Average([\mathbf{h}_{L-k+1}^{u}, \mathbf{h}_{L-k+2}^{u},\cdots, \mathbf{h}_{L}^{u}]).
\end{equation}
Similarly, we use the same strategy to obtain the aggregated representation $\mathbf{h}_{u}^{+}$ for $\vmathbb{S}_u^{+}$.
Then we adopt InfoNCE loss to pull the positive instances closer and push negative examples away in the semantic space, which can be represented as:
\begin{equation}
\begin{aligned}
\mathcal{L}_{\text {ICL}}=\sum_{u=1}^{|\mathcal{U}|}-\log \frac{e^{\operatorname{sim}\left(\mathbf{h}_{u}, \mathbf{h}_{u}^{+}\right) / \tau}}{e^{\operatorname{sim}\left(\mathbf{h}_{u}, \mathbf{h}_{u}^{+}\right) / \tau} + \sum_{i\neq u, i \in M(u)} e^{\operatorname{sim}\left(\mathbf{h}_{u}, \mathbf{h}_{i}\right) / \tau}},
\end{aligned} \label{con:contrastiveequa}
\end{equation}
where $M(u)$ denotes the users in the mini batch that contains $u$, $sim(\cdot, \cdot)$ represents cosine similarity function, and $\tau$ is the temperature.
\subsection{Learning Equivariance for Invasive Augmentation}
\label{invasive}
A straightforward solution for predicting invasive augmentation is to use a simple linear layer as the prediction head $\phi_{\zeta}(\cdot)$ in Equation 3, which may be sub-optimal when the corrupted sequence is easily reconstructable.
Based on our objective to improve the models' ability to learn equivariance, we adopt a generator-discriminator structure inspired by ELECTRA~\cite{clark2020electra} to obtain hard invasively augmented views. 
The generator's goal is to produce the “hard” positive sequences that exhibit minor differences compared to the original ones, while the discriminator aims to accurately detect even the slightest changes introduced by the generator. To this end, the Replaced Item Detection Loss (RIDL) is adopted as the prediction loss on masked item substitution to facilitate the equivariance learning.
Below is an example of how to learn equivariance for item substitution with the generator-discriminator structure:

Given a user behavior sequence $\vmathbb{S}_u=[v^u_1, v^u_2, \dots, v^u_{L}]$, we first randomly mask several items of $\vmathbb{S}_u$ with the mask ratio $\gamma$. The masked interaction sequence can be represented as $\vmathbb{S}_{u}^{\prime}= m^{u}\cdot \vmathbb{S}_u$, where $m^{u}=[m^u_1, m^u_2, \dots, m^u_{L}]$ and $m^u_{t}\in\{0,1\}$.
Then we use BERT4Rec~\cite{bert4rec} as a generator $f_{\mathcal{G}}(\cdot)$ to recover randomly masked items in $\vmathbb{S}_{u}^{\prime}$ to obtain the partially substituted user interaction sequence $\vmathbb{S}_{u}^{\prime\prime}=f_{\mathcal{G}}(\vmathbb{S}_{u}^{\prime})$.
With this method, we construct fake interaction sequences with smaller semantic perturbations, which poses a challenge for the discriminator to correctly identify the substituted items.
The generator is trained with the following loss:
\begin{equation}
\mathcal{L}_{Gen}=\sum_{u=1}^{|\mathcal{U}|}\sum_{v_{m} \in \vmathbb{S}_{u}^{m}}-\frac{1}{\left|\vmathbb{S}_{u}^{m}\right|} \log p\left(v_{m}=v_{m}^{*} \mid \vmathbb{S}_{u}^{\prime}\right),
\end{equation}
where $\vmathbb{S}_{u}^{m}$ is the set of masked items in $\vmathbb{S}_{u}^{\prime}$, $v_{m}^{*}$ denotes the ground truth item for the masked item $v_{m}$, and the $p(\cdot)$ is the same function as Equation~\ref{eq:rec_loss}.

The Conditional Discriminator (CD) aims to perform the Replaced Item Detection (RID) task regarding the output representations of the UBE $\mathbf{h}_{t}^{u} = f_{\theta}(\{v^{u}_{j}\}_{j=1}^{t})$ as conditions.
In this way, the gradients of CD propagate back to the UBE,  which encourages the UBE to generate more informative user behavior representations, so that CD could distinguish the tiny discrepancy between $\vmathbb{S}_u$ and $\vmathbb{S}_{u}^{\prime\prime}$. 
In our implementation, we use the SASRec~\cite{sasrec} blocks and an extra MLP layer to instantiate the CD (denoted as $f_{\mathcal{D}}(\cdot)$).
We use aggregation function such as concatenation to inject  conditional information into the discriminate modeling process.
For each item in the user sequence, the CD needs to predict whether it has been substituted or not. 
We calculate the cross-entropy loss as follows:

\aptLtoX[graphic=no,type=html]{
\begin{equation} 
\label{eq:dis-bce}
\begin{split} \mathcal {L}_{RID} = \sum _{u=1}^{|\mathcal {U}|} \sum _{t=2}^{L} &- \vmathbb{1}(v_{t}^{u^{\prime \prime }}=v_{t}^{u}) \log \sigma (\mathbf {w}^{T}f_{\mathcal {D}}(\vmathbb {S}_{u}^{\prime \prime },\mathbf {h}_{t}^{u},t)) \\ &- \vmathbb{1}(v_{t}^{u^{\prime \prime }}\ne v_{t}^{u}) \log (1-\sigma (\mathbf {w}^{T}f_{\mathcal {D}}(\vmathbb {S}_{u}^{\prime \prime },\mathbf {h}_{t}^{u},t))), \end{split} 
\end{equation} }{
\begin{equation}
\label{eq:dis-bce}
\begin{split}
\mathcal{L}_{RID} = \sum_{u=1}^{|\mathcal{U}|} \sum_{t=2}^{L} 
&- \mathbbm{1}(v_{t}^{u^{\prime\prime}}=v_{t}^{u})
\log \sigma(\mathbf{w}^{T}f_{\mathcal{D}}(\vmathbb{S}_{u}^{\prime\prime},\mathbf{h}_{t}^{u},t)) \\ 
&- \mathbbm{1}(v_{t}^{u^{\prime\prime}}\neq v_{t}^{u})
\log (1-\sigma(\mathbf{w}^{T}f_{\mathcal{D}}(\vmathbb{S}_{u}^{\prime\prime},\mathbf{h}_{t}^{u},t))),
\end{split}
\end{equation}}
where $\mathbf{w}$ is a learnable parameter matrix, $\sigma$ denotes the sigmoid function and $v_{t}^{u^{\prime\prime}} \in \vmathbb{S}_{u}^{\prime\prime}$.

\noindent \textbf{Why the proposed ECL-SR can achieve equivariance?} 
As described in Sec.~\ref{ECL}, equivariance is the property that when a group transformation $\mathcal{T}_g$ is applied to the input sequence $\vmathbb{S}_u$, the resulting output features also undergo a corresponding transformation $\mathcal{T}_g^{\prime}$, which can be mathematically represented as  $f_\theta\left(\mathcal{T}_g(\boldsymbol{\vmathbb{S}_u})\right)=\mathcal{T}_g^{\prime}(f_\theta(\boldsymbol{\vmathbb{S}_u}))$ where $f_\theta$ is the encoder and $\mathcal{T}_g$ is a transformation from group $A$. 
In ECL-SR, equivariance is achieved because the designed generator and conditional discriminator can encourage the user behavior encoder to detect the semantic changes (namely the difference between $\mathcal{T}_g(\boldsymbol{\vmathbb{S}_u})$ and original sequence $\vmathbb{S}_{u}$) caused by invasive augmentations rather than ignoring them, which has been theoretically supported by recent work in computer vision~\cite{ecl}\footnote{Readers can check Proposition 1 and its proof in~\cite{ecl}.}.
It should be noted that our proposed ECL-SR differs from previous works on equivariance~\cite{cohen2016group, bronstein2021geometric}, as it only encourages equivariant properties through the choice of the loss function RIDL, rather than enforcing strict equivariance.
To this end, ECL-SR treats each type of invasive augmentation as a group and uses a conditional discriminator to predict the presence of augmentations (\textit{e.g.}, item substitution) in the input sequence. 

\subsection{Optimization}
\subsubsection{Model Training and Inference}
In the training stage, item embeddings are shared across all three modules. The parameters of the user behavior encoder and discriminator (except for the extra linear layer for RID) are also shared to avoid overfitting.
All components of the ECL-SR model are trained in an end-to-end manner.
Therefore, the whole ECL-SR framework is optimized with the combined loss function:
\begin{equation}
\label{eq:all_loss}
	\mathcal{L}=\mathcal{L}_{\text {Rec}}+\lambda_{1}\mathcal{L}_{\text {ICL}}+\lambda_{2}\mathcal{L}_{\text {Gen}}+\lambda_{3} \mathcal{L}_{\text {RID}},
\end{equation}
where $\lambda_{\cdot}$ controls the contribution of each auxiliary loss.
In the inference stage, we remove both the generator and conditional discriminator, and only use the UBE to complete the next item prediction task.
\subsubsection{Model Complexity}
\label{sec:complexity}
The complexity of the instantiated ECL-SR stems from 3 parts:  user behavior encoder (UBE), generator (G) and conditional discriminator (CD).
All of them share the same embedding table, which contains the majority of the parameters.
The complexity of UBE+CD is close to that of SASRec~\cite{sasrec}, as their parameters are shared to enhance training stability and efficiency. 
The complexity of G remains close to that of BERT4Rec~\cite{bert4rec}. 
Consequently, the overall complexity of the ECL-SR is comparable to that of SASRec combined with BERT4Rec (sharing embedding table).
 To ensure computational efficiency similar to that of SASRec, we maintain the same total number of layers in UBE and G as in other self-attentive methods used in the experiments.
Furthermore, to minimize computational overhead, we fix the parameters of G after training for 10 epochs. 
During inference, the speed of ECL-SR is comparable to SASRec as only the UBE is utilized. 
We summarize the model complexity comparison in Tab.~\ref{tab:complexity}.

\begin{table}
    \small
	\caption{Statistics of the datasets after pre-processing.}
	\label{tab:datasets}
	\setlength{\tabcolsep}{0.6mm}{
	\begin{tabular}{lrrrr}
	\toprule
	Dataset   & Sports & Toys & Yelp & ML-1m  \\
	\midrule
	\# Users  & 35,598 & 19,412 & 30,499 & 6,041\\
	\# Items  & 18,357 & 11,924 & 20,068 & 3,417 \\
	\# Avg. Actions / User  & 8.3 & 8.6 & 10.4 & 165.5\\
	\# Avg. Actions / Item  & 16.1 & 14.1 & 15.8 & 292.6 \\
	\# Actions  & 296,337 & 167,597 & 317,182 & 999,611\\
	Sparsity  & 99.95\% & 99.93\% & 99.95\% & 95.15\%\\
	\bottomrule
	\end{tabular}
	}
\end{table}

\section{Experiment}
In this section, we conduct comprehensive experiments to answer the
following key research questions:
\begin{itemize}
    \item \textbf{RQ1:} 
    Can ECL-SR achieve competitive performance compared with state-of-the-art basic SR methods and ICL-based SR methods?
    \item \textbf{RQ2:} How do different types of invasive or mild data augmentation affect the performance of ECL-SR?
    \item \textbf{RQ3:} How does each module in ECL-SR contribute to its performance?
    \item \textbf{RQ4:} What is the effect of hyper-parameters in ECL-SR?
\end{itemize}

\subsection{Experimental Settings}
\subsubsection{Dataset.}
We conduct experiments on 4 widely used benchmark datasets with diverse distributions: \textbf{Sports} and \textbf{Toys} is constructed from Amazon review datasets\footnote{http://jmcauley.ucsd.edu/data/amazon/}~\cite{DBLP:conf/sigir/McAuleyTSH15};
\textbf{Yelp}\footnote{https://www.yelp.com/dataset} is a famous business recommendation dataset;
\textbf{ML-1m}\footnote{https://grouplens.org/datasets/movielens/1m/} is a famous movie rating dataset comprising 1 million ratings.
We pre-process these datasets in the same way following ~\cite{sasrec,zhou2020s3,yuan2021icai,tang2018personalized,yuan2019simple,coserec} by removing items and users that occur less than 5 times. Tab.~\ref{tab:datasets} shows dataset statistics after pre-processing.

\subsubsection{Evaluation Metrics.}
Following previous works ~\cite{bert4rec, Xie2022DIF, coserec, ICLRec}, we use two widely adopted metrics to evaluate the performance of SR models: Recall@$K$ and top-$K$ Normalized Discounted Cumulative Gain (NDCG@$K$), where $K$ is selected from from $\{10,20\}$. 
For each user's interaction sequence, we reserve the last two items for validation and test, respectively, and use the rest to train SR models. 
As suggested in ~\cite{krichene2020sampled,dallmann2021case}, we calculate the scores for each item in the entire item set and then sort them in descending order, aiming to ensure a more fair comparison among items.
\subsubsection{Baseline Models.}
\label{baselines}
We compare our proposed ECL-SR with 12 different baselines, including 2 general methods, 5 strong basic SR methods, and 5 state-of-the-art invariant contrastive learning based SR methods. The introduction of these baselines are as follows:
\begin{itemize}
    \item \textbf{PopRec}: A straightforward method where items are ranked based on popularity. Specifically, items with higher popularity, i.e., those interacted with by more users, are more likely to be recommended to users.
    \item \textbf{BPR}~\cite{bpr}: A classic method employing pairwise comparisons and matrix factorization to optimize the ranking of items based on user preferences.. 
    \item \textbf{Caser}~\cite{tang2018personalized}: A CNN-based method leveraging both horizontal and vertical perspective for personalized sequential recommendation.
    \item \textbf{GRU4Rec}~\cite{hidasi2015session}: A RNN-based method that adopts Gated Recurrent Units (GRUs) with the ranking loss for session-based recommendation.
    \item \textbf{SASRec}~\cite{sasrec}: A self-attentive SR model that utilizes a unidirectional Transformer to capture users' dynamic preferences.
    \item \textbf{BERT4Rec}~\cite{bert4rec}: A bidirectional Transformer-based SR model that adopts the cloze task for training.
    \item \textbf{ELECRec}~\cite{elecrec}: The first SR model to train sequential recommenders as discriminators instead of generators.
    \item \textbf{S$^{3}$Rec$_{MIP}$}~\cite{zhou2020s3}: A self-supervised SR method that devises four different pretext tasks to improve the quality of item representations. For a fair comparison, we only use Masked Item Prediction (MIP) as the pre-training task following previous papers~\cite{coserec,duorec}
    \item \textbf{CL4SRec}~\cite{cls4rec}: An invariant contrastive learning-based SR model that employs three sequence-level augmentation operators to generate positive pairs.
    \item \textbf{CoSeRec}~\cite{coserec}: 
    An invariant contrastive learning-based SR model that utilizes item correlations to generate high-quality positive pairs. To this end, two informative data augmentation strategies are devised.
    \item \textbf{DuoRec}\footnote{ECL-SR is a type of unsupervised contrastive method, so for a fair comparison, we only use the Unsupervised Contrastive Learning (UCL) variant of DuoRec proposed in their original paper for a fair comparison.}~\cite{duorec}: An invariant contrastive learning-based SR model that obtains contrastive samples through feature-level dropout masking and the supervised positive sampling.
    \item \textbf{ICLRec}~\cite{ICLRec}: A general learning paradigm that leverages the clustered latent intent factor and contrastive self-supervised learning to optimize SR.
\end{itemize}
\subsubsection{Implementation Details.}
To ensure a fair comparison, we implement all baselines and ECL-SR using the popular recommendation framework RecBole~\cite{zhao2021recbole} and conduct evaluations under the same settings. We train the models using the Adam optimizer for 200 epochs, with a batch size of 256 and a learning rate of 1e-4. Other hyper-parameters are set as follows: the maximum sequence length is 50 for Sports, Toys, and Yelp, and 200 for ML-1m; the layer number and head number are both set to 2 for ECL-SR and attention-based baselines; the hidden size is set to $256$; the dropout rate on the embedding matrix and attention matrix is set to 0.5 for Sports and Toys, and 0.2 for ML-1m and Yelp; the temperature $\tau$ in the contrastive loss is set to 0.05; the masking ratio in the generator is set to 0.2 for Sports, Toys, and ML-1m, and 0.6 for Yelp; the window size $k$ is set to 5 for Sports, Toys, and Yelp, and 20 for ML-1m; the balance parameter $\lambda_{\cdot}$ is set to 0.1 for the replaced item detection loss $\mathcal{L}_{\text {RID}}$, 0.2 for the generator loss $\mathcal{L}_{\text {Gen}}$, and 0.3 for $\mathcal{L}_{\text {ICL}}$.
\begin{table*}[t]
  \caption{Overall performance. 
  The best results are highlighted in bold. Scores with an underline are ranked second. "*" indicates that there is a statistically significant difference (p<0.01) when compared to the best baseline methods using paired t-tests.
}
  \label{tab:overall}
  \resizebox{.975\textwidth}{!}{\setlength{\tabcolsep}{0.85mm}{
      \begin{tabular}{l|cccc|cccc|cccc|cccc}
        \toprule
        \multicolumn{1}{c|}{\multirow{3}{*}{SR Model}} & 
        \multicolumn{4}{c|}{\multirow{1}{*}{Sports}} &
        \multicolumn{4}{c|}{\multirow{1}{*}{Toys}} &
        \multicolumn{4}{c|}{\multirow{1}{*}{Yelp}} &
         \multicolumn{4}{c}{\multirow{1}{*}{ML-1m}} 
          \\
          \cline{2-17}
          & 
        \multicolumn{2}{c}{\multirow{1}{*}{Recall}}&
        \multicolumn{2}{c|}{\multirow{1}{*}{NDCG}}
         &
        \multicolumn{2}{c}{\multirow{1}{*}{Recall}}&
        \multicolumn{2}{c|}{\multirow{1}{*}{NDCG}}
        &  
        \multicolumn{2}{c}{\multirow{1}{*}{Recall}}&
        \multicolumn{2}{c|}{\multirow{1}{*}{NDCG}} &
        \multicolumn{2}{c}{\multirow{1}{*}{Recall}}&
        \multicolumn{2}{c}{\multirow{1}{*}{NDCG}}
        \\
         &  
        @10 & @20 &
        @10 & @20 &
        @10 & @20 &
        @10 & @20 &
        @10 & @20 &
        @10 & @20 &
        @10 & @20 &
        @10 & @20 \\
        \midrule
        PopRec & 0.0146 &  0.0244 & 0.0078 & 0.0103 & 0.0105 & 0.0172 &  0.0060 &  0.0077 & 0.0099& 0.0161& 0.0051& 0.0067&  0.0204&  0.0417 & 0.0101 &  0.0154\\
        BPR & 0.0302 &0.0480 & 0.0144 &0.0188 & 0.0344& 0.0560 &0.0151&0.0205 & 0.0589&0.0830&0.0324&0.0384 & 0.0260&0.0553&0.0115&0.0188 \\
        \hline
        GRU4Rec & 0.0386 & 0.0609 & 0.0195 & 0.0251 & 0.0449 & 0.0708 & 0.0221 & 0.0287 & 0.0418 & 0.0679 & 0.0206 & 0.0271 & 0.1781 & 0.2725 & 0.0886 & 0.1123  \\
        
        Caser & 0.0227&0.0364&0.0118&0.0153 & 0.0361&0.0566&0.0186&0.0238 & 0.0380&0.0608&0.0197&0.0255 & 0.1579&0.2515&0.0624&0.0992\\
        
        BERT4Rec & 0.0295&0.0465&0.0135&0.0173 & 0.0533&0.0787&0.0234&0.0297 & 0.0524&0.0756&0.0327&0.0385 & 0.1275& 0.2370&0.0584&0.0852\\
        
        SASRec & 0.0511 & 0.0781 & 0.0226 & 0.0294 & 0.0796  & 0.1181  & 0.0360  & 0.0457 & 0.0667 &  0.0969 & 0.0405 & 0.0481 & 0.2145 & 0.3228 & 0.1074 & 0.1346  \\
        ELECRec & 0.0480 & 0.0758 & 0.0234 & 0.0301 & 0.0784 & 0.1134 & 0.0356 & 0.0444 & 0.0677 & 0.0980 & 0.0406 & 0.0482 & 0.1917 & 0.2917  & 0.0902 & 0.1154  \\
        \hline
        S$^3\text{Rec}_{\text{MIP}}$ & 0.0509 & \underline{0.0787} & 0.0236 & 0.0306 & 0.0792& 0.1157 & 0.0362 & 0.0455 & 0.0637 & 0.0929 & 0.0392 & 0.0465 & 0.1992 & 0.3086 & 0.0961 & 0.1237 \\

        CL4SRec & 0.0497 & 0.0782 & \underline{0.0248} & \underline{0.0320} & 0.0808 & 0.1162  & 0.0372 & 0.0461 & 0.0640 & 0.0937 & 0.0387 & 0.0462 & 0.2200 & 0.3341 & 0.1120 & 0.1408  \\
        CoSeRec & 0.0514 & 0.0778 & 0.0223 & 0.0299 &  0.0789 & 0.1144  & 0.0368 & 0.0458 & 0.0648& 0.0947 & 0.0379 & 0.0454 & 0.2214 & 0.3210 & 0.1032 &  0.1285 \\

        DuoRec & 0.0479 & 0.0708 & 0.0246 & 0.0304 & 0.0805& 0.1136 & 0.0363 & 0.0460 & 0.0635& 0.0982 & 0.0336 & 0.0423 & \underline{0.2392} & 0.3336  & \underline{0.1271} & \underline{0.1509}  \\
        ICLRec & \underline{0.0527} & 0.0773 & 0.0236 & 0.0298 & \underline{0.0826}& \underline{0.1183} & \underline{0.0377} & \underline{0.0467} & \underline{0.0701}& \underline{0.1005} & \underline{0.0422} & \underline{0.0499} & 0.2291 & \underline{0.3445}  & 0.1140 & 0.1431  \\
        \hline
        \textbf{ECL-SR} & \textbf{0.0547*} &\textbf{0.0835*} &\textbf{0.0272*} & \textbf{0.0343*} & \textbf{0.0843*} & \textbf{0.1202*} & \textbf{0.0379*} & \textbf{0.0468*} & \textbf{0.0739*} & \textbf{0.1059*} & \textbf{0.0430*} & \textbf{0.0506*} &\textbf{0.2563*} & \textbf{0.3692*} & \textbf{0.1311*} & \textbf{0.1593*} \\
        
      \bottomrule
    \end{tabular}
}
}
\end{table*}
\subsection{Overall Performance (RQ1)}
Tab. \ref{tab:overall} presents the overall performance of ECL-SR and other baselines on all datasets.
Based on the empirical results, we have the following observations. First, the SR methods significantly exceed general recommendation baselines, including PopRec and BPR, demonstrating the importance of leveraging sequential patterns.
For the 5 basic SR models, ELECRec and SASRec beat other methods with a large improvement, while BERT4Rec achieves better or close performance than GRU4Rec and Caser in most cases. This demonstrates the advantage of attention-based methods in processing sequential data. Note that the recently proposed discriminator-based model ELECRec does not show significant performance gain on most datasets if we carefully tune SASRec. This suggests that directly applying a discriminator on the next item prediction (NIP) task may not be an optimal choice for SR.

For 5 invariant contrastive learning based SR methods, we notice that they may not consistently outperform the fully-tuned SASRec. This suggests that applying invasive augmentation directly to invariant contrastive learning may not always be beneficial and could potentially be detrimental in certain situations. Among these methods, ICLRec demonstrates superior performance in the majority of cases. This can be attributed to using learned intent prototypes as positive representation of a given sequence, which to some extent mitigates the semantic shift caused by invasive augmentation. These improvements also highlight the need for more effective utilization of invasive augmentation strategies.

Finally, ECL-SR consistently beats both state-of-the-art basic SR models and contrastive learning-based SR models across all datasets and evaluation metrics.
Specifically, on Yelp and ML-1m, ECL-SR has \textbf{10.8\%} and \textbf{19.5\%} respective \nobreak improvements over vanilla SASRec in terms of Recall@10. It also has \textbf{5.4\%} and \textbf{7.1\%} improvement in Recall@10 compared to the strongest invariant contrastive learning baseline. 
This consistent improvement demonstrates that our proposed solution effectively helps make next-item predictions on both sparse and dense datasets. 
ECL-SR can adaptively encourage the learned user behavior representation to be equivariant to invasive augmentation while keeping invariant to other mild augmentation via optimizing contrastive loss and replaced item detection loss at the same time. 
This sophisticated design can fully utilize self-supervision signals to guide the model in capturing behavior patterns hidden in user-item interaction data.
Additionally, as shown in Tab.~\ref{tab:complexity}, our proposed ECL-SR model does not significantly increase the number of training parameters, training time, and inference time when compared to state-of-the-art contrastive learning based SR models. This can be attributed to the strategies discussed in Section~\ref{sec:complexity} for reducing computational complexity.
\begin{table}[t]
\centering
\caption{Model complexity of ECL-SR and ICL-based models on Yelp dataset.}
\label{tab:complexity}
\resizebox{75mm}{!}{%
\begin{tabular}{c|ccc}
\hline
Models & \# Params. & Train time per epoch & Inference speed \\ \hline
DuoRec & 1.3877M & 160s & 1846.48/s \\
ELECRec & 1.3880M & 165s & 1866.04/s \\
CLS4Rec & 1.3877M & 156s & 1850.75/s \\
CoSeRec & 1.3877M & 161s & 1840.19/s \\
ECL-SR & 1.3995M & 163s & 1827.60/s \\ \hline
\end{tabular}%
}
\end{table}
\subsection{Impact of Augmentation Types (RQ2)}
\label{rq2}
As mentioned in Sec.~\ref{sec-intro}, there exist several options for mild augmentation and invasive augmentation, leading to different combinations of augmentation strategies. This section investigates their impacts on the performance of ECL-SR .
\subsubsection{Impact of Different Mild Augmentation Methods}
We experiment with dropout, feature normalization, and feature perturbation as mild augmentation in ECL-SR.
The details of feature normalization and perturbation are shown in Sec.~\ref{sec-aug}.
For a fair comparison, we fix the invasive augmentation, i.e. masked item substitution, and use grid search to obtain the best performances of ECL-SR with respect to these three different mild augmentations on ML-1m and Yelp. 
The results are shown in Tab.~\ref{tab: mild-augment}.
We can observe that for all datasets, dropout achieves better performance than both normalization and perturbation.

\subsubsection{Impact of Different Invasive Augmentation Methods}
We test masked item substitution, random item substitution, insertion, deletion, crop and reorder as invasive augmentation in ECL-SR. For a fair comparison, we fix the mild augmentation method to be dropout and search the optimal hyper-parameter for every invasive augmentation. Tab.~\ref{tab: mild-augment} shows the performance of the invasive augmentation methods, and we can see that masked item substitution significantly outperforms the other five invasive augmentation methods in most cases.
In particular, masked item substitution brings more performance gain compared with random item substitution. A potential reason may be that masked item substitution (achieved by BERT4Rec~\cite{bert4rec}) better preserves pivotal correlations among items in the original user sequence, thus empowering the conditional discriminator to capture small differences.

\begin{table*}[t]
\centering
\caption{Performance comparison with respect to different mild and invasive augmentations.}
\resizebox{130mm}{!}{
\begin{tabular}{l|c c|c c|c c|c c}
\toprule
\multirow{2}*{\textbf{Settings}} &  \multicolumn{2}{c|}{\textbf{ML-1m}} & \multicolumn{2}{c|}{\textbf{Yelp}}  & \multicolumn{2}{c|}{\textbf{Toys}} & \multicolumn{2}{c}{\textbf{Sports}} \\
~ & Recall@10 & NDCG@10 & Recall@10 & NDCG@10 & Recall@10 & NDCG@10& Recall@10 & NDCG@10\\ 
\midrule
\multicolumn{1}{l}{\textit{Mild Augmentation}}\\
\hline
Dropout &  \textbf{0.2563} & \textbf{0.1311} & \textbf{0.0739} & \textbf{0.0430} & \textbf{0.0843} & \textbf{0.0379} & \textbf{0.0547}  & \textbf{0.0272}   \\ 
Norm & 0.2502 & 0.1276 & 0.0730 & 0.0426 & 0.0827 & 0.0376  & 0.0526 & 0.0254\\ 
Perturb &  0.2520& 0.1295 & 0.0715 & 0.0424 & 0.0818 & 0.0371 & 0.0524 & 0.0264\\ 
\hline
\multicolumn{1}{l}{\textit{Invasive Augmentation}}\\
\hline
Substitution (Mask)& \textbf{0.2563} & \textbf{0.1311} & \textbf{0.0739} & \textbf{0.0430} & 0.0843 & \textbf{0.0379}  & \textbf{0.0547}  & \textbf{0.0272} \\ 
Substitution (Random)& 0.2474 & 0.1265 & 0.0705 & 0.0417 & 0.0828 & 0.0374 & 0.0503 & 0.0247\\ 
Insert  &  0.2435 & 0.1271 & 0.0680 & 0.0413 & 0.0808 & 0.0372  & 0.0528 & 0.0265 \\ 
Delete & 0.2454 & 0.1277 & 0.0707 & 0.0416 & \textbf{0.0848} & 0.0384  & 0.0505 & 0.0253\\
Crop & 0.2464 & 0.1255 & 0.0701& 0.0416 & 0.0843 & 0.0379  & 0.0515 & 0.0256\\
Reorder & 0.2460 & 0.1273 & 0.0694 & 0.0418 & 0.0827 & 0.0378 & 0.0497 & 0.0248\\
\bottomrule
\end{tabular}}
\label{tab: mild-augment}
\end{table*}

\subsection{Ablation Study (RQ3)}
\begin{figure*}
    \centering
    \includegraphics[width=130mm]{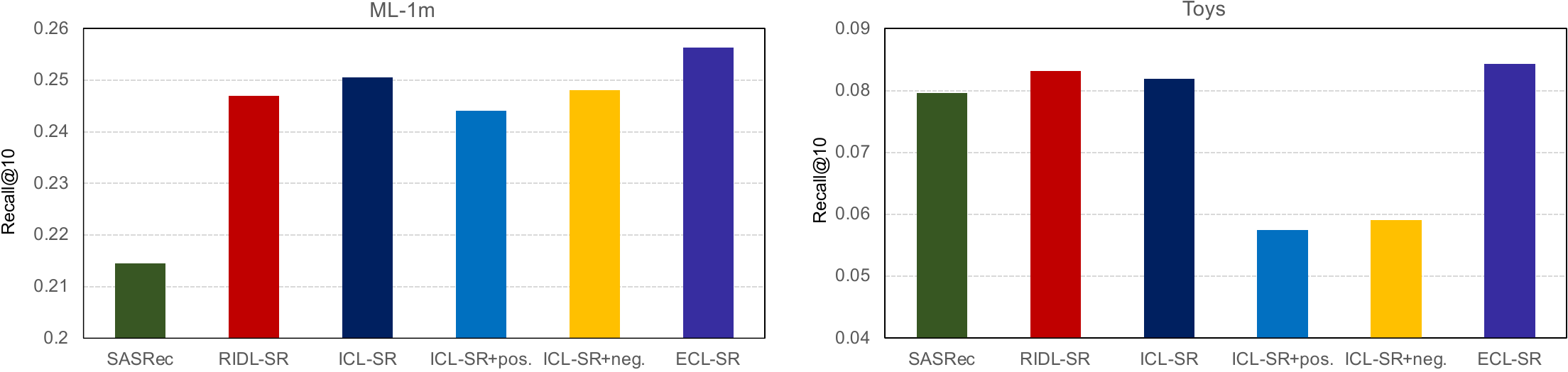}
    \caption{Ablation study of ECL-SR on ML-1m and Toys datasets.}
    \label{fig:ablation-cl}
\end{figure*}

To investigate the contribution of the different components and to support the model design, we introduce the following ECL-SR variants for the ablation study.
\begin{itemize}
    \item \textit{ECL-SR w/o contrastive loss \& replaced item detection doss (RIDL) }, which is identical to \textbf{SASRec}.
    \item \textit{ECL-SR w/o contrastive loss}, which only learns equivariance for invasive augmentation through RIDL\footnote{
We do not explore learning equivariance for mild augmentation because RIDL cannot be directly applied to "mild" augmentations: "mild" augmentations work at the feature level while RIDL detects the augmentations at item level.}. For brevity, we denote this variant as \textbf{RIDL-SR}.
    \item \textit{ECL-SR w/o RIDL}, which stops detecting the differences caused by invasive augmentation and only uses contrastive loss for mild augmentation. We also call this variant as Invariant Contrastive Learning for SR (\textbf{ICL-SR}).
    \item \textit{\textbf{ICL-SR + pos}.}, an extension of ICL-SR that regards the augmented sequences recovered by the generator (e.g., pretrained BERT4Rec) as additional positive intances for the contrastive loss in Eq.~\ref{con:contrastiveequa}.
    \item \textit{\textbf{ICL-SR + neg.}},  an extension of ICL-SR that regards the augmented sequences recovered by the generator (e.g., pretrained BERT4Rec) as additional negative intances for the contrastive loss in Eq.~\ref{con:contrastiveequa}.
\end{itemize}
We summarize the ablation results on Toys and ML-1m in Fig.~\ref{fig:ablation-cl}.
By comparing \textit{SASRec}, \textit{RIDL-SR} and \textit{ICL-SR}, we can see that both contrastive loss and replaced item detection loss contribute to performance improvement. This observation verifies that both invariance and equivariance are important properties for enhancing the quality of item representation. The recommendation performance is further enhanced by the combination of CL and RIDL in ECL-SR, demonstrating the effectiveness of leveraging the complementary nature between invariance and equivariance in representation learning.

A natural concern for ECL-SR is whether the generator, which generates augmented sequences by replacing masked items, is the primary contributor to the observed performance enhancements.
To figure this out, we first pretrain the generator and then feed the augmented sequences produced by the pre-trained generator into the contrastive loss for training the ICL-SR variant. 
We conduct separate experiments for the augmented sequences used as positive and negative samples, denoted as \textit{ICL-SR+pos.} and \textit{ICL-SR+neg.} and the results are presented in Fig.~\ref{fig:ablation-cl}.
We observe that using a pre-trained generator to produce positive or negative instances for contrastive loss calculation weakens the recommendation performance on the Toys and ML-1m datasets, compared to ICL-SR that only uses dropout masking to generate positive instances.
This observation is consistent with the empirical results shown in Fig~\ref{fig:intro}, which shows that mild augmentations are more suitable than invasive augmentations for invariant contrastive learning.
It also demonstrates the need to encourage user behavior representation to be equivariant to invasive data augmentation by introducing an additional pretext task (e.g., replaced item detection).

\subsection{Hyper-parameter Sensitivity Study (RQ4)}
This section examines the impact of 5 critical hyper-parameters in ECL-SR, including batch size, the temperature $\tau$ in Equation~\ref{con:contrastiveequa}, window size $k$, mask ratio $\gamma$, and dropout ratio.
We adjust one hyper-parameter at a time and keep the others optimal to control variables.
Due to space constraints, we only put the results of Yelp and Toys in Fig.\ref{fig:hyper}, as the results of other two datasets exhibit similar trends for most hyper-parameters.
\subsubsection{Impact of Batch Sizes.}
In ECL-SR, training batch size determines the number of negative instances/views for contrastive learning because, for each user sequence, all other user sequences in the batch are regarded as negative instances.
Intuitively, the more negative instances we get, the higher the items' quality is.
With higher-quality item representations, the SR model could predict the next item more accurately.
However, as shown in Fig~\ref{fig:hyper} (a), ECL-SR achieves the best performance when the batch size is 256, and larger batch sizes do not lead to better performance. The reason for this could be attributed to the fact that when the batch size gets larger, there are more false negatives in each batch.

\subsubsection{Impact of Temperatures $\tau$.}
Previous studies~\cite{wang2021understanding,you2022momentum} show that the temperature $\tau$ plays an important role in determining the strength of penalties on hard negative samples due to global uniformity and local separation of embedding distributions.
In particular, contrastive loss with extremely small $\tau$ tends to concentrate only on the nearest one or two samples, which may disrupt the formation of useful features for the recommendation task. 
In contrast, an extremely large $\tau$ can result in a lack of uniformity when generating embeddings.
Therefore, $\tau$ should be kept in an appropriate interval to ensure the effectiveness of contrastive learning.
From Fig~\ref{fig:hyper} (b), we observe that for Toys and Yelp, 0.1 and 0.05 are the optimal values of $\tau$, respectively.
\subsubsection{Impact of Window Sizes $k$.}
\label{rq3}
This experiment aims to investigate the effect of the window size $k$ of the last-$k$ aggregation strategy in Sec~\ref{mild}. We report the results in Fig \ref{fig:hyper} (c). 
 We observe that $k$ of 10 and 5 achieve the best performance on Toys and Yelp, respectively.
Both smaller or larger value lead to performance degradation, indicating a trade-off between local patterns and distant patterns in user behavior modeling.
\subsubsection{Impact of Mask Ratios $\gamma$}
The mask ratio $\gamma$ is used to control how many items would be substituted in the original sequence for RID task, which is described in Sec.\ref{invasive}. 
A higher $\gamma$ means more perturbation in the sequence and vice versa. We visualize the Recall@10 scores with respect to different values of $\gamma$ in Fig~\ref{fig:hyper} (d).
We observe that for Yelp and Toys dataset, the performance of ECL-SR fluctuates slightly when $\gamma$ falls between 0.2 and 0.5, and the best performance is achieved when $\gamma$ is set to 0.6 and 0.2, respectively.
\subsubsection{Impact of Dropout Ratios.}
\begin{figure*}
    \centering
    \includegraphics[width=\textwidth]{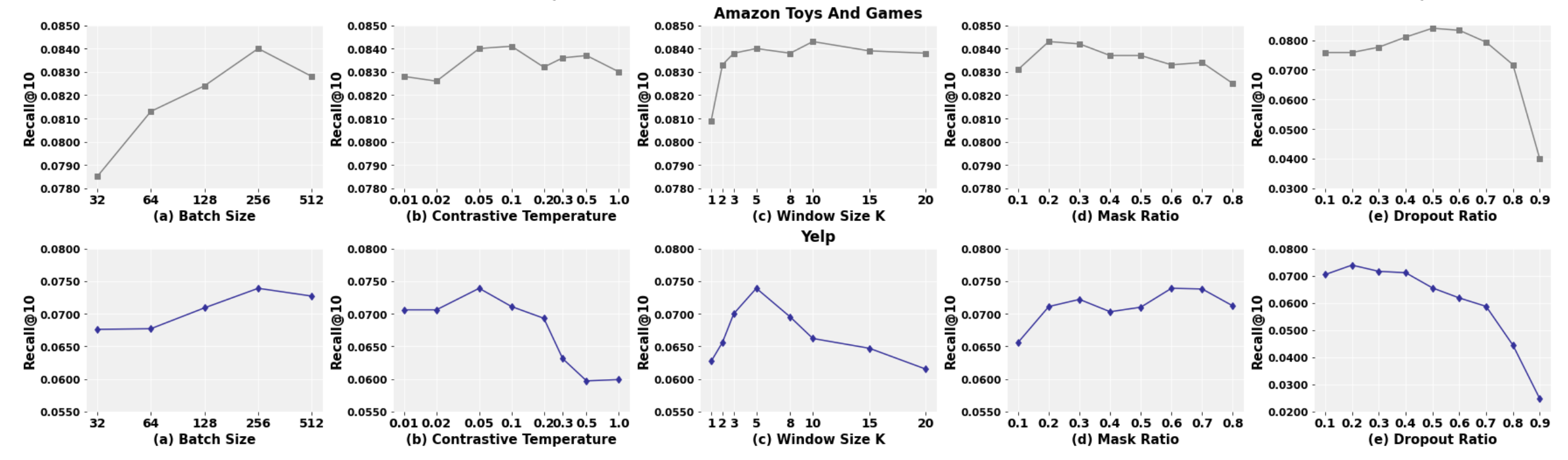}
    \caption{Performance (Recall@10) comparison with respect to 5 hyper-parameters on Toys and Yelp datasets. }
    \label{fig:hyper}
\end{figure*}

The dropout ratio regulates how much we deflect the semantics of the original user sequence when generating positive views.
As we can observe in Fig~\ref{fig:hyper} (e), the optimal dropout ratio for Toys and Yelp are 0.5 and 0.2, respectively.
We also find that when we increase the dropout ratio to large values, such as 0.9, the model performance decreases substantially on both datasets.
We attribute this to the fact that extremely large dropout ratios typically make the semantics of positive views much different from the original sequence, and maximizing the consistency between them largely reduces the quality of item representations.
\section{Related Work}
\subsection{Sequential Recommendation}
The Sequential Recommendation System (SRS) aims to discover users' preferences and interests from their historical behavior and predict the next items they would most like to interact with~\cite{wang2022veracity,wang2022trustworthy,xie2023rethinking}. Essentially, prediction accuracy depends largely on SRS's ability to identify complex dependencies within user interactions.
In the early days, traditional data mining and machine learning methods, such as pattern mining~\cite{yap2012effective} and Markov chains~\cite{he2016fusing,rendle2010factorizing}, were used to model simple low-order sequential dependencies. However, these methods rely on rigid assumptions and cannot handle complex data~\cite{wang2023data}. 
During the past few years, deep learning networks, such as recurrent neural networks (RNN)~\cite{gru4rec,hidasi2018recurrent,quadrana2017personalizing}, convolutional neural networks (CNN)~\cite{tang2018personalized,yuan2019simple}, memory networks~\cite{rum,huang2018improving}, and graph neural networks (GNN)~\cite{chang2021sequential,ma2020memory,zhang2022efficiently}, have gained popularity for sequential recommendations due to their strong representational capabilities. 
Recently, transformer-based models~\cite{sasrec,bert4rec,elecrec} have received increasing attention because of their ability to model long- and short-term item dependencies in sequences.
Typically, SASRec~\cite{sasrec} uses self-attention mechanism to dynamically adjust the weights associated with items that users have previously interacted with, achieving excellent performance on both sparse and dense datasets.
To encode users' dynamic preferences more effectively, BERT4Rec~\cite{bert4rec} proposes to use a deep bidirectional self-attention model to capture both left and right-side contextual information.
In the follow-up studies, multiple Transformer variants were developed to adapt to different scenarios through integrating side information~\cite{Xie2022DIF,liu2023chatgpt}, time intervals~\cite{li2020time}, personalization~\cite{personalization}, additional local constraints~\cite{he2021locker}, and reducing model complexity~\cite{lightweight}.  
However, all the above methods are designed to model transition patterns in a supervised learning manner, which does not take into account the sparsity problem caused by limited user observations.
In this paper, we propose alleviating this data sparsity problem by introducing an equivariant contrastive learning paradigm in the architecture of Transformers.

\subsection{Contrastive Learning}
In the past few years, Contrastive Learning (CL) has achieved great success in various research areas, including computer vision ~\cite{kang2020contragan}, natural language processing ~\cite{simcse}, speech processing ~\cite{saeed2021contrastive,chong2023masked}, and more.
The main purpose of CL is to produce high-quality and information-rich representations by reducing the distance between positive views generated from the same data instance and separating their negative views in the latent space.
There have also been some recent efforts to apply contrastive learning to sequential recommendations.
For instance,
S$^{3}$Rec~\cite{zhou2020s3} designs four auxiliary self-supervised objectives to enhance data representations by maximizing mutual information among attributes, items, subsequences, and sequences.
CL4SRec~\cite{cls4rec} employs three data-level augmentation operators (crop/mask/reorder, referred to as invasive augmentation methods in this paper) to construct positive views.
Later, CoSeRec~\cite{coserec} proposes to generate robust augmented sequences based on item correlation since random item perturbations may undermine the confidence of positive pairs.
DuoRec~\cite{duorec} leverages dropout-based feature-level augmentation (referred to as mild augmentation in this paper) to better maintain semantic consistency between positive views, which alleviates representation degeneration in CL4SRec.
ICLRec~\cite{ICLRec} uses clustering techniques to obtain the user's intent distribution from all user behavior sequences and optimizes the SR model by comparing the sequence with corresponding intentions.

Unlike the above approaches, our work is the first to apply equivariant contrastive learning to sequential recommendation, which can improve the semantic quality of user behavior by encouraging models to learn both invariance and equivalence from mild augmentation and invasive augmentation.

\section{Conclusion}
In this paper, we propose the ECL-SR framework, which effectively utilizes both mild and invasive augmentation to enhance user behavior representations.
Specifically, we introduce the use of a conditional discriminator to capture the user behavior discrepancy between the original interaction sequence and its edited version, which has been shown to be a useful objective for encouraging the user behavior encoder to be equivariant to masked item substitution augmentation.
Our experiments on four benchmark SR datasets demonstrate the effectiveness of ECL-SR, which achieves competitive performance compared with both classical SR models and invariant contrastive learning based SR models. In the future, we plan to explore more combinations of data augmentation methods using ECL-SR framework.

\bibliographystyle{ACM-Reference-Format}
\bibliography{sample-base}


\end{document}